\documentclass[aps,prl,twocolumn,showpacs,superscriptaddress,floatfix]{revtex4}
\usepackage{epsfig}

\newcommand{\be}{\begin{equation}}
\newcommand{\bea}{\begin{eqnarray}}
\newcommand{\eea}{\end{eqnarray}}
\newcommand{\ee}{\end{equation}}

\newcommand{\bra}[1]{\mbox{$\langle #1 |$}}
\newcommand{\ket}[1]{\mbox{$| #1 \rangle$}}

\begin{document}

\title{Quantum Information Processing in Strongly Detuned Optical Cavities.}

\author{E. Jan\'e}
\affiliation{Departament d'Estructura i Constituents de la Mat\`eria, Universitat de Barcelona, E-08028 Barcelona, Spain}
\affiliation{QOLS, Blackett Laboratory, Imperial College of Science, Technology and Medicine, London, SW7 2BW, UK}
\author{M.B. Plenio}
\affiliation{QOLS, Blackett Laboratory, Imperial College of
Science, Technology and Medicine, London, SW7 2BW, UK}
\author{D. Jonathan}
\affiliation{DAMTP, Centre for Mathematical Sciences, University of Cambridge, Wilberforce Road, Cambridge CB3 0WA, UK}

\date{\today}

\begin{abstract}
We demonstrate that quantum information processing can be
implemented with ions trapped in a far detuned optical cavity. For
sufficiently large detuning the system becomes insensitive to
cavity decay. Following recent experimental progress, this scheme can be
implemented with currently available technology.
\end{abstract}
 \maketitle
Quantum information and quantum entanglement are the key resources
in quantum information science \cite{Plenio V 98}. Achieving their
controlled experimental manipulation is of paramount importance
for the actual implementation of any quantum information
processing protocol. Many schemes for the practical implementation
of quantum information processing have been proposed in the past.
Amongst those, one of the most promising approaches towards
controlled and scalable quantum information processing employs
linear ion traps. They offer the possibility in principle to store
strings of ions, cool them almost to zero temperature and address
them individually with laser light \cite{Experiments}. The
individual ions can interact with each other via their motional
degrees of freedom. Indeed, it was realized early on that, based
on these features, ion traps allow for the implementation of all
quantum gates that are necessary for universal quantum computation
\cite{Cirac Z 95}. In practice however, the implementation of this
proposal is less straightforward. Very low temperatures have to be
achieved and subsequent heating of the ion motion as well as
spontaneous emission from the ions has to be suppressed
\cite{decoherence}. Many proposals such as heat-insensitive
quantum gates \cite{Sorensen M 98,Jonathan P 01,Schneider JM 00}
and the use of Raman transitions and interference-aided laser
cooling \cite{Morigi EK 00} have been made to address these
problems. Nevertheless, the requirement of achieving low
temperatures in the motional degrees of freedom of the ions proves
to be a significant problem in actual experiments.

Recently, approaches have therefore been proposed which employ
ions that are trapped inside of an optical cavity. Instead of
using the mechanical degrees of freedom of the ions to mediate
their interaction, here the optical cavity mode plays that role.
The inevitable cavity decay through the mirrors appears to cause a
problem. However, it has been realized that in fact this cavity
decay may be used to create entanglement between ions in cavities
\cite{Plenio HBK 99} and to establish quantum communication
networks between different optical cavities \cite{Bose KPV 99}.

In this paper we also study a system comprising trapped ions inside
an optical cavity, a system that has been demonstrated
experimentally in various groups \cite{Lange+Schmidt-Kaler}. The
cavity mirrors are not perfect, but as opposed to earlier
proposals we do not employ this cavity decay but rather use the
idea that a strongly detuned cavity can be used to suppress cavity
decay. In fact, the use of strongly detuned cavities will
effectively decohere all those transition amplitudes that would
lead to a population of the cavity mode, thereby suppressing the
decay of the cavity. The spontaneous decay of the internal
electronic states of the ions can also be reduced by using ions in
a configuration where two ground states representing the qubit are
coupled via far detuned Raman transitions.

In the following we will demonstrate that with this set-up
comprising trapped ions and a far detuned cavity we are able to
implement both single qubit rotations as well as two qubit gates
such as the CNOT gate. Therefore universal quantum computation can
be achieved in our system. We also present a physical picture,
based on the dynamical Stark shifts, to provide insight into the
physical origin of the function of the gates.

\begin{figure}[htb]
\epsfysize=4.5cm
\begin{center}
\epsffile{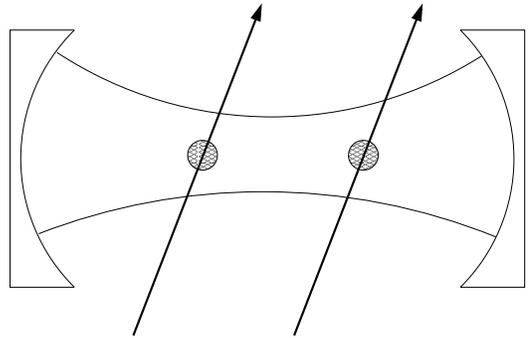}
\end{center}
\caption{\label{figure_cavity} We consider a set-up in which two
ions are trapped inside of a strongly detuned optical cavity.
For our scheme to work, the ions have to separated by at least
one optical wavelength.}
\end{figure}

\begin{figure}[htb]
\epsfysize=4.cm
\begin{center}
\epsffile{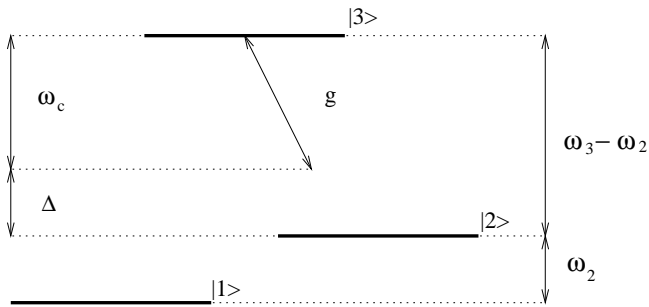}
\end{center}
\caption{\label{figure_ion} Energy diagram of one of the ions. The cavity
drives the transition $\ket{2}\rightarrow\ket{3}$ with coupling constant $g$
and has a detuning $\Delta$.}
\end{figure}

Consider the situation in which there are two ions in a cavity as
shown in figure \ref{figure_cavity}. We assume that the ions are
separated by at least one optical wavelength. Both ions have the
internal structure of a lambda system and the optical cavity
couples the states $\ket{2}$ and $\ket{3}$ of each ion. The lower
levels, states \{$\ket{1}_i$,$\ket{2}_i$\} ($i=1,2$ denotes the
ion), form our qubit, while $\ket{3}$ is an auxiliary state. The
energy level structure for one of the ions is shown in figure
\ref{figure_ion}. The Hamilton operator describing the combined
cavity-ion system can be written as
\bea\label{hamiltonian_1}
    H &=& \omega_2\left(\sigma_{22}^1+\sigma_{22}^2\right)+\omega_3
    \left(\sigma_{33}^1+\sigma_{33}^2\right)+\omega_c a a^\dagger
    \nonumber\\
    && + g\{a^\dagger\left(\sigma^1_{23}+\sigma^2_{23}\right)+h.c.\},
\eea
where $\sigma^i_{\alpha\beta}=\ket{\alpha}_{ii}\bra{\beta}$, $g$
is the cavity-ion coupling constant, $a$ and $a^\dagger$ are the
annihilation and creation operators for the cavity photons and
$\hbar=1$.

As mentioned in the introduction, the cavity is strongly detuned,
more precisely, we assume, that the detuning
$\Delta=\omega_c-(\omega_3-\omega_2)$ and the cavity-atom coupling
constant satisfies the condition $g^2/\Delta^2\ll 1$. A
consequence of the strong detuning of the cavity is that the
population in the cavity mode will be very small at all times if
initially it is initially not populated. As states involving
photons play a negligible role in the time evolution we can
eliminate them from the Hamilton operator adiabatically
\cite{James ..}. To do this we compute the effective Hamiltonian
(in an interaction picture with respect to
$H_0=\sum_i\left(\omega_2\sigma_{22}^i+\omega_3
    \sigma_{33}^i\right)+(\omega_3-w_2) a a^\dagger$) to second order for the subspace
with zero photons, which yields

 \bea\label{hamiltonian_2}
    H_{\mathrm{eff}}&=&-\frac{g^2}{\Delta}(\sigma^1_{11}\sigma^2_{33}
    +\sigma^1_{22}\sigma^2_{33}+\sigma^1_{33}\sigma^2_{11}+
    \sigma^1_{33} \sigma^2_{22}\nonumber
    \\ &&\phantom{-\frac{g^2}{\Delta}(}+2\sigma^1_{33}\sigma^2_{33}+\sigma^1_{32}
    \sigma^2_{23}+ \sigma^1_{23}\sigma^2_{32}).
\eea
All contributions that could induce transitions to levels
involving photons are rapidly oscillating due to the strong
detuning of the cavity mode.

Furthermore, from eq. (\ref{hamiltonian_2}) we see that the
detuned cavity induces an energy shift in the energy-levels of the
atoms. The result is that the state
\be
    \ket{\Psi_a}=\frac{1}{\sqrt{2}}(\ket{23}-\ket{32}),
\ee
together with the states
\{$\ket{11}$,$\ket{12}$,$\ket{21}$,$\ket{22}$\} are not shifted
whereas the rest are shifted by an amount of order
$\frac{g^2}{\Delta}$. The goal will be to use these energy shifts
to create a conditional dynamics of the two qubits.

Now we drive the transition $\ket{2}_i\rightarrow\ket{3}_i$ in
both ions with two lasers with Rabi frequency $\Omega_1=\Omega$
for ion $1$ and $\Omega_2=-\Omega$ for ion $2$. The full
Hamiltonian of the system in the interaction picture, taking into
account eq. (\ref{hamiltonian_2}), is
\bea\label{hamiltonian_3}
    H'&=&H_1+H_2\nonumber\\
    H_1&=&-\frac{g^2}{\Delta}(\sigma^1_{11}\sigma^2_{33}+\sigma^1_{22}
    \sigma^2_{33}+\sigma^1_{33}\sigma^2_{11}+\sigma^1_{33}\sigma^2_{22}
    \nonumber \\
    &&\phantom{-\frac{g^2}{\Delta}(}+2\sigma^1_{33}\sigma^2_{33}
    +\sigma^1_{32}\sigma^2_{23}+\sigma^1_{23}\sigma^2_{32})\nonumber\\
    H_2&=&\Omega \left(\sigma^1_{32}-\sigma^2_{32}+h.c.\right).
\eea

If the condition $(g^2/\Delta)^2\gg \Omega^2$ is satisfied, we can
again use perturbation theory \cite{perturbation} to find the
effective Hamiltonian for the subspace where the states that we
want to use as qubits belong to. From eq. (\ref{hamiltonian_4})
one can see the this subspace is
\{$\ket{11}$,$\ket{12}$,$\ket{21}$,$\ket{22}$,$\ket{\Psi_a}$\},
since the rest of the states are shifted out of resonance by the
interaction of the ions with the cavity. The effective Hamiltonian
to first order gives
\be\label{hamiltonian_4}
    H''=\Omega\sqrt{2}\left(\ket{22}\bra{\Psi_a}+\ket{\Psi_a}\bra{22}\right).
\ee
Note that because some of the levels are shifted, the laser only
couples the state $\ket{22}$ with the state $\ket{\Psi_a}$. The
resulting effective Hamiltonian is thus non-local despite the
local interaction of the laser with the ions. The explanation for
this apparent contradiction lies in the non-local character of the
ion-cavity interaction. The Hamiltonian eq. (\ref{hamiltonian_4})
can be employed to perform the CNOT gate because the
transformation
\bea\label{gate1}
    &&\ket{11}\rightarrow\ket{11}\nonumber\\
    &&\ket{12}\rightarrow\ket{12}\nonumber\\
    &&\ket{21}\rightarrow\ket{21}\nonumber\\
    &&\ket{22}\rightarrow-\ket{22},
\eea
can be obtained via $e^{-iHt}$ after a time
$t=\frac{\pi}{\sqrt{2}}\Omega^{-1}$, since after a complete
oscillation between the state $\ket{\Psi_a}$ and $\ket{22}$ the latter
gains a phase, whereas the rest of the states do not evolve. This
control-phase gate can be turned into a CNOT gate by applying
before and after it a Hadamard transformation.

\begin{figure}[bht]
\epsfysize=3cm
\begin{center}
\epsffile{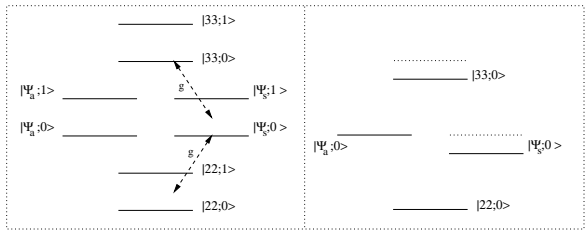}
\end{center}
\caption{\label{energylevels} Selected energy levels of the ions
for both zero and one photon levels. Due to the detuned
atom-cavity interaction (indicated by arrows) energy shifts occur
in some of the zero-photon levels.  }
\end{figure}

The physical idea behind this scheme is presented in figure
\ref{energylevels} where we depict the ionic energy levels
corresponding to zero and one photons in the cavity mode. The
off-resonant coupling between the ion and the cavity (indicated by
arrows) induce Stark-shifts in some of the energy levels
corresponding to zero photon states. Choosing a suitably detuned
laser, we can therefore selectively couple energy levels. As the
energy shifts arise due to the interaction with a cavity mode
which is non-local we obtain an effective non-local dynamics when
shining lasers on individual ions.
This idea is reminiscent to the use of free-space dipole-dipole
interaction between extremely closely spaced ions
\cite{dipoledipole} but here the interaction is mediated by the
cavity mode. It should be noted, that in our set-up the ions can
have a distance of many optical wavelengths at which the free
space dipole-dipole interaction is negligible.

\begin{figure}[hbt]
\epsfysize=6cm
\begin{center}
\epsffile{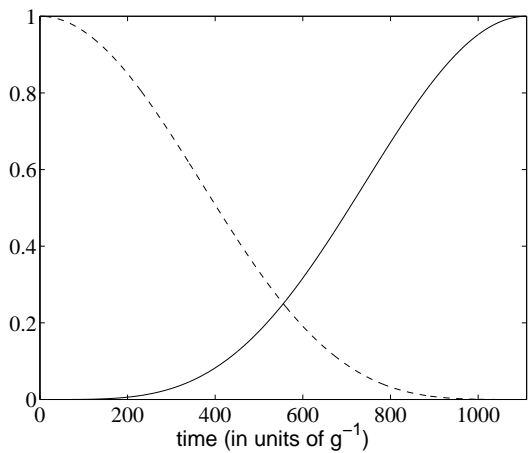}
\end{center}
\caption{\label{figure_plot1} Oscillations between the states
$1/\sqrt{2}(\ket{11}+\ket{22})$ and
$1/\sqrt{2}(\ket{11}-\ket{22})$ when there is no dissipation. In
the figure one can see the populations of both states. The
parameters are $\Delta=3 g$ and $\Omega=2.0\,10^{-3}g$. The probability
of having zero photons is $>0.999$.}
\end{figure}

In order to see whether the above approximations are correct, we
have studied numerically the system using the full Hamiltonian
without any approximations. For instance, we would expect
oscillations between the state $1/\sqrt{2}(\ket{22}+\ket{11})$ and
$1/\sqrt{2}(\ket{22}-\ket{11})$. In figure \ref{figure_plot1} we
plot the oscillations between these two states for the parameters
$\Delta=3.0 g$ and $\Omega=0.01 g$. We observe that although
$(g/\Delta)^2\sim 1$, the photon population remains vanishingly
small for all practical purposes. It is important to note that in
this scheme the interaction of the cavity-mode with the two ions
is the ingredient that allows us to construct the two-qubit gate
although the cavity mode has (almost) no population throughout the
performance of the gate. This becomes important when the cavity
has losses since in principle this will not affect this procedure.

\begin{figure}[ht]
\epsfysize=6cm
\begin{center}
\epsffile{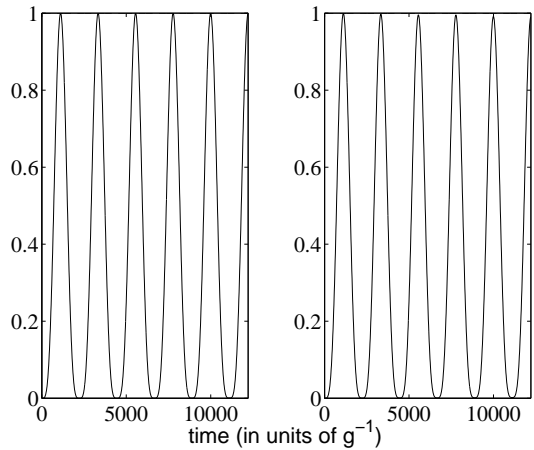}
\end{center}
\caption{\label{fidelity} Fidelity of the gate starting from the
state $1/\sqrt{2}(\ket{11}+\ket{22})$ without cavity decay, ie
$\kappa=0$ (left) and with cavity decay $\kappa=1.0 g$ (right). The
gate is performed $5$ consecutive times. The parameters are
$\Delta=3.0 g$ and  $\Omega=2.0\,10^{-3}g $. The similarity of
the two graphs demonstrates that the influence of the cavity decay
is efficiently suppressed. }
\end{figure}

It remains to be seen that any one-qubit gate can be performed
within this set-up, since single qubit operations are required to
achieve universal quantum computation. A general single qubit
rotation can be achieved by coupling the two levels $\ket{1}$ and
$\ket{2}$ of one ion via a Raman transition using two lasers with
the appropriate phases. This transition can be done using another
level which is not coupled to the cavity and does not suffer any
energy shift. This framework allows us to work with states with a
very long lifetime and therefore the decoherence due to
spontaneous emission of the ions when they are in the lower levels
is strongly suppressed.

In the analysis so far we have neglected the cavity decay. In a
practical experimental situation, however, such process is present
and for practical purposes, is important to know whether the
scheme is robust against decoherence. Fortunately, it turns out
that the use of a far detuned cavity makes the scheme robust
against cavity decay as the cavity modes have an extremely small
population throughout the gate operation. This is demonstrated in
figure \ref{fidelity} where we have run a simulation with
vanishing $\kappa$ (left half of the graph) and $\kappa=g$ (right
half of the graph. There is no appreciable difference between the
two time evolutions which clearly shows that the influence of the
cavity decay is efficiently suppressed.

\begin{figure}[ht]
\epsfysize=6cm
\begin{center}
\epsffile{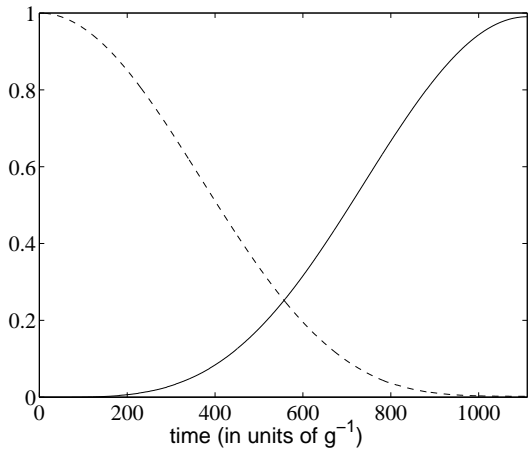}
\end{center}
\caption{\label{figure_mc} Oscillations between the states
$1/\sqrt{2}(\ket{11}+\ket{22})$ and
$1/\sqrt{2}(\ket{11}-\ket{22})$. In the figure one can see the
populations of both states. The parameters are $\Delta=3.0 g$,
$\Omega=2.0\,10^{-3}g$, the cavity decay rate $\kappa=0.5 g$ and
$\Gamma=5.0\,10^{-5}g$ for the line-width of the upper level.}
\end{figure}

A further source of decoherence in this proposal is the fact that
the state $\ket{3}$ can decay spontaneously to $\ket{2}$ since the
upper level is populated during the evolution. On the other hand
the gate can be performed very fast because even when we have
$g\sim\Delta$ there are no photons created in the cavity. In
figure \ref{figure_mc} we show a Monte Carlo simulation
\cite{QJumps} where we see that we have a gate fidelity of $\sim
0.99$ for a decay rate of $\Gamma=5.0\cdot 10^{-4}g$.

In summary, we have demonstrated that a system consisting of ions
trapped inside of a far detuned optical cavity can be used to
implement coherent quantum information processing. The strong
detuning between cavity mode and ion leads to an effective
suppression of the cavity decay. Spontaneous decay from the ions
is reduced by storing the qubits in two lower states of a Raman
configuration. Both single qubit and two-qubit can be implemented.
The scheme can be implemented with current experimental technology
as the recent preparation of trapped ions in optical cavities
demonstrates \cite{Lange+Schmidt-Kaler}.

We would like to thank J. Pachos for discussions and for bringing
his closely related results \cite{Pachos W01} to our attention
before publication. This work was partially supported by EPSRC,
the European Union EQUIP project the ESF QIT program and grants
MEC (AP99), IUAP-P4-02 and GOA-Mefisto-666 .



\begin{thebibliography}{99}
\bibitem{perturbation} In order to be able to do perturbation
theory the condition $\Omega\gg g^4/\Delta^3$ has to be satisfied,
since (\ref{hamiltonian_2}) has corrections of order ${\cal
O}(g^4/\Delta^4)$.
\end{thebibliography}
\end{document}